\documentclass[aps, prx, reprint]{revtex4-2}

\usepackage{amsfonts}
\usepackage{amsmath}
\usepackage{amssymb, bm, mathrsfs}
\usepackage{graphicx, epstopdf, color}
\usepackage{soul}
\usepackage{braket}
\usepackage[colorlinks=true,citecolor=blue]{hyperref}

\begin{document}

\title{Altermagnetic Schottky Contact}

\author{Yee Sin Ang}
\thanks{Author to whom correspondence should be addressed: yeesin\_ang@sutd.edu.sg}
\affiliation{Science and Math Cluster, Singapore University of Technology and Design (SUTD), 8 Somapah Road, Singapore 487372.}

\begin{abstract}

Altermagnet is an emerging antiferromagnetic material subclass that exhibits spin-splitting in momentum space without net global magnetization and spin-orbit-coupling effect. In this work, we develop a model of thermal charge injection across an altermagnet/semiconductor (AM/S) Schottky contact. We obtain analytical expressions describing the spin-dependent thermionic current injection across the AM/S contact under any arbitrary interface orientation angles. 
Interestingly, the spin-contrasting Fermi surface of an altermagnetic electrode enables spin-polarized current to be injected into a nonmagnetic semiconductor even though the system has net-zero magnetization. Our findings thus reveal an altermagntic mechanism to achieve spin injection without involving ferromagnetic ordering. The AM/S Schottky contact proposed here shall provide a potential altermagnetic building block of spintronic devices that are resilient against stray magnetic field perturbation and inherently compatible with ultracompact integration.

\end{abstract}

\maketitle

Metal/semiconductor (MS) contact \cite{10.1063/1.4858400} is a ubiquitous building block of modern electronic and optoelectronic device technology. MS contact can serve as a standalone device, offering myriads of useful device applications covering diodes \cite{di2016graphene}, energy converters \cite{zhang2020tribovoltaic, jie2013graphene}, solid-state refrigeration \cite{mahan1994thermionic, o2005electronic, o2007thermionic, zhang2023thomson} and sensors \cite{zhao2020reversible}, as well as forming a key component of electronics and optoelectornics devices, such as field-effect transistor (FET) \cite{liu2021promises} and light emitting diode (LED) \cite{malliaras1998roles}. Taking FET as an example, the metallic electrode contacting the semiconducting channel forms an MS contact across which the charge injection efficiency critically influences the energy consumption and the switching speed of the device \cite{quhe2021sub}.
Beyond conventional transistor device architecture, \emph{magnetic} MS interface composed of ferromagnetic (FM) metal and semiconductor \cite{ohno1999electrical} allows the injection of spin-polarized current into a nonmagnetic semiconductor \cite{hammar1999observation, fert2001conditions, albrecht2003spin, smith2008spin, chen2013control, gan2013two},  thus offering a key enabler of spintronic device technology \cite{datta1990electronic}. 
MS contact and the charge injection physics across MS contact \cite{ang2021physics, zheng2021ohmic} is thus of foundational importance for designing a wide array of electronics, optoelectronics \cite{xu2016contacts, lv2021design} and spintronics devices \cite{schmidt2005concepts}. 

Altermagnet is an emerging class of materials with antiferromagnetic ordering, but exhibit finite spin-splitting in the momentum space even when the system does not posses spin-orbit coupling (SOC) effect \cite{PhysRevX.12.031042, PhysRevX.12.040501, PhysRevB.102.014422, vsmejkal2020crystal}. 
Altermagnet can host spin polarized transport \cite{PhysRevLett.119.187204} and spin Hall effect \cite{zhang2018spin} without SOC, thus circumventing the need of complex crystal structures and heavy elements that are commonplace to generate strong SOC \cite{PhysRevB.102.014422}.
The absence of macroscopic net magnetization also minimizes stray field effect that are especially problematic in high-density device application \cite{bai2020functional}. 
Because of the highly anisotropic nature of the spin-split band structure in altermagnets, directional-dependent and spin-polarized transport arises in both normal \cite{shao2021spin, das2023transport, PhysRevLett.128.197201}, ferromagnetic \cite{sun2023spin} and superconducting tunneling heterostructures \cite{PhysRevB.108.054511, PhysRevB.108.L060508, PhysRevLett.131.076003, zhang2023finite} as well as the bulk transport \cite{PhysRevLett.130.216701, huang2022antiferromagnetic, PhysRevLett.126.127701}, thus enabling the altermagnetism to be sensitively probed via transport measurement. 
A large variety of altermagnet has been discovered, ranging from 2D and 3D crystals \cite{PhysRevX.12.040501} to insulating \cite{noda2016momentum} and metallic \cite{PhysRevB.99.184432} phases. Recent high-throughput study based on the MAGNDATA database \cite{gallego2016magndata, gallego2016magndata2} uncovers 60 species of altermagnets \cite{guo2023spin}, which further broadens the application potential of altermagnetic materials. 
Importantly, the antiferromagnetic ordering and the Neel vector of altermagnet can be electrical switched by via spin-orbit torque (SOT) \cite{takeuchi2021chiral}, thus leading to myriads of potential spintronic device applications \cite{shao2021spin, PhysRevLett.128.197201}.

\begin{figure*}[t]
    \includegraphics[scale=0.5]{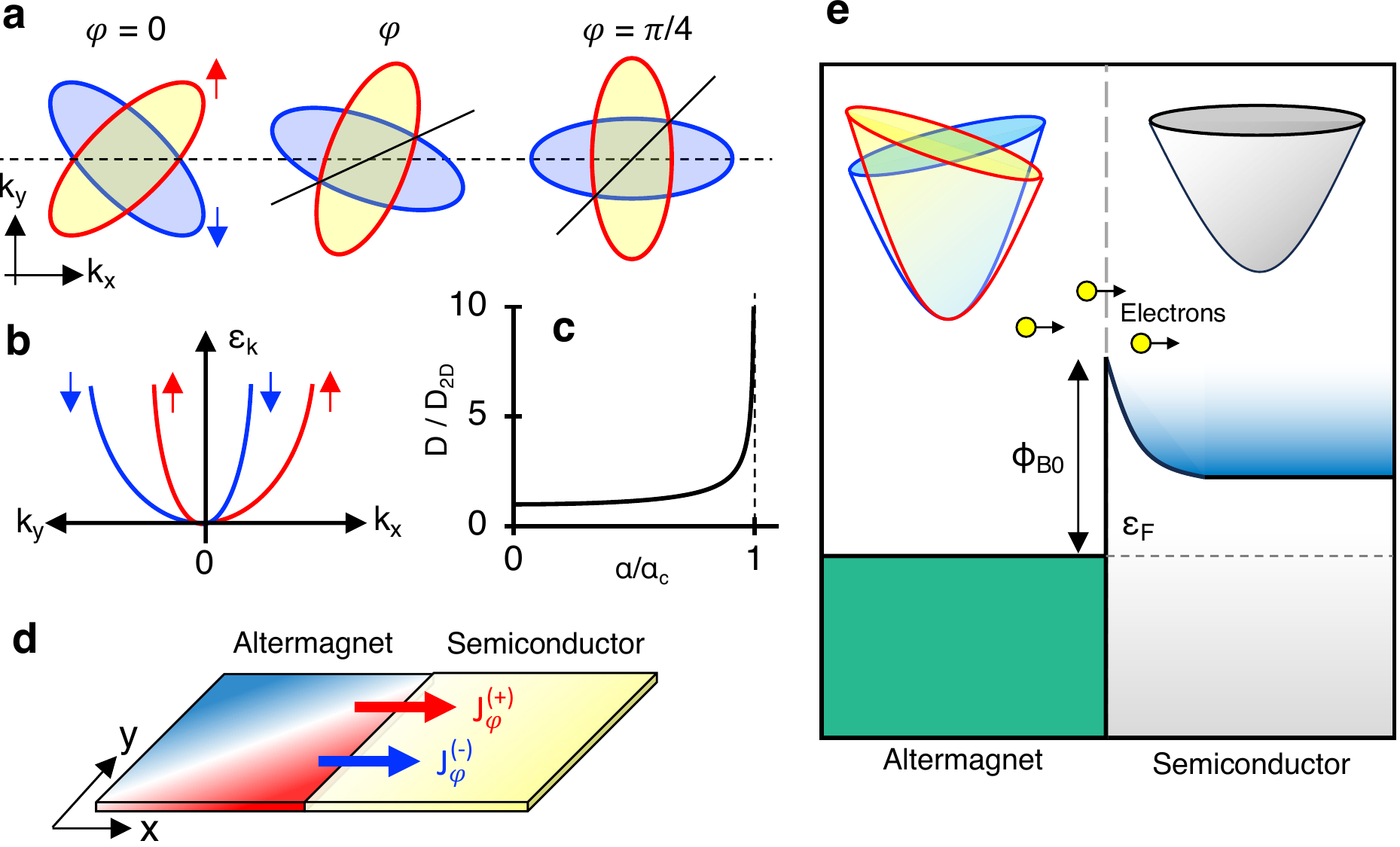}
    \caption{\textbf{Electronic properties of altermagnet and concept of altermagnet/semiconductor Schottky contact.} (a) Fermi surfaces of altermagnet at various orientation axis $\varphi$. (b) Energy dispersion along $k_x$ and $k_y$ directions. (c) Electronic density of states (normalized by that of 2DEG) as a function of altermagnetic strength. (d) Schematic drawing of an altermagnet/semiconductor Schottky contact. (e) Band diagram of the altermagnet/semconductor Schottky contact.}
    \label{fig:1}
\end{figure*}

In this work, we attempt to address a straightforward yet potentially technologically important question that arises when the concepts of MS contact and altermagnet are put together: Can altermagnets provide a route to achieve spin-polarized current injection into a nonmagnetic semiconductor via an \emph{altermagnetic} MS Schottky contact -- counterpart to that composed of ferromagnetic electrodes? 
By constructing a theory of thermal charge injection across an altermagnetic-metal/nonmagnetic-semiconductor Schottky contact, we show that altermagnetic Schottky contact can host substantial spin-polarized transport even though SOC and a global magnetization are both absent. 
The injection current exhibit strong interface orientation dependence due to the anisotropic spin-split Fermi surface.
The analytical expression developed in this work reveals the existence of a maximal universal spin polarization efficiency of 26.8\% in an altermagnetic Schottky contact. 
Although such spin polarization efficiency does not outperform the ferromagnetic counterparts, the presence of a sizable spin polarization efficiency is still an unexpected results considering that both Zeeman and Rahsba-Dresselhaus interactions are completely absent in an altermagnetic Schottky contact. 
These findings generalize the concept of Schottky contact into the altermagnetic realm, and demonstrate the momentum-dependent spin polarization in altermagnets as an alternative pathway to inject spin-polarized current into a nonmagnetic semiconductor without involving any ferromagnetic ordering and SOC. 
The absence of stray field and SOC suggest that altermagnetic Schottky contact may provide a building block of spintronic devices that are inherently compatible with ultracompact setup and resilient against external magnetic field. 

We consider the two-band effective Hamiltonian of a metallic altermagnet that have been previously employed to study the normal and superconductor tunneling junctions \cite{PhysRevB.108.054511, PhysRevB.108.L060508}, $\hat{\mathcal{H}}_{\mathbf{k}} = \alpha_c k^2 \sigma_0 + \alpha k_x k_y \sigma_z$ where $\alpha_c \equiv \hbar^2/2m^*$, $m^*$ is the electron effective mass, $\alpha$ is a parameter describing the strength of altermagnetic spin-plitting, $\sigma_0$ is the $2\times2$ identity matrix, $\sigma_z$ is the $z$-component Pauli spin matrix, $\mathbf{k} = (k_x,k_y)$ is electron wave vector and $k = \left|\mathbf{k}\right|$. To model the interface orientation, we rotate the altermagnet Hamiltonian by an angle of $\varphi$ with respect to the $x$-axis, yielding
\begin{equation}
    \hat{\mathcal{H}}_{\textbf{k},\varphi} = \alpha_c k^2 \sigma_0 + \alpha \left[\psi(\varphi) \left(k_x^2 - k_y^2\right) - k_xk_y \psi'(\varphi)\right] \sigma_z
\end{equation}
where $\psi(\varphi) \equiv \sin\varphi\cos\varphi$ and $\psi'(\varphi) \equiv \sin^2\varphi - \cos^2\varphi$. The energy dispersion is:
\begin{equation}
    \varepsilon_{\mathbf{k}, \varphi}^{(s)} = \Lambda_+^{(s)}(\varphi) k_x^2 + \Lambda_-^{(s)}(\varphi) k_y^2 - s\alpha \psi'(\varphi) k_xk_y
\end{equation}
where $s=\pm1$ denotes the spin-up and spin-down subbands, $\Lambda_\pm^{(s)} \equiv \alpha_c \pm s\alpha \psi$. In the following, we suppress the argument for simplicity: $\psi(\varphi) \to \psi$, $\psi'(\varphi) \to \psi'$ and $\Lambda_\pm^{(s)}(\varphi) \to \Lambda_\pm^{(s)}$. The Fermi surface is elliptical and exhibits spin-contrasting orientation in the momentum space [Fig. \ref{fig:1}(a)]. While the $\varphi = 0$ orientation does not generate spin-polarized transport along the $x$-direction, the $\varphi = \pi/4$ orientation, in which the major (minor) axis of the spin-up (spin-down) Fermi surface lobe aligns the $x$-axis [see Fig. \ref{fig:1}(b) for the energy dispersion versus $k_x$ and $k_y$], can host a spin-polarized $x$-directional transport current. An analytical expression of the total electronic density of states (DOS) can be conveniently obtained from the energy dispersion at $\varphi = \phi/4$ via $D(\varepsilon_k) = \sum_s \int \delta\left( \varepsilon_{\mathbf{k}, \pi/4}^{(s)} - \varepsilon_k\right) \text{d}^2k$ where $\delta(x)$ is the Dirac delta function, which yields 
\begin{equation}
    D(\varepsilon_k) = D_\text{2D} \left[1 - \left(\frac{\alpha}{\alpha_c}\right)^2 \right]^{-1/2}
\end{equation}
where $D_\text{2D} = m/\pi \hbar^2$ is the DOS of a 2DEG \cite{sze2021physics}. Here the DOS is singular as $\alpha/\alpha_c \to 1$ [see Fig. \ref{fig:1}(c)]. Such singularity arises from the energy dispersion which becomes hyperbolic as $\alpha > \alpha_c$ and we focus on the case of $\alpha< \alpha_c$ in the following.

The group velocity along the $x$ direction is
\begin{equation}\label{vg}
    v_x^{(s)} (k_x, k_y) = \hbar^{-1} \frac{\partial \varepsilon_{\mathbf{k},\varphi}^{(s)}}{ \partial k_x} = 2\Lambda_+^{(s)}k_x^2  - s\alpha \psi' k_y.
\end{equation}
Considering a contact heterostructure composed of altermagnetic metal and a nonmagnetic semiconductor [Fig. \ref{fig:1}(d)], the contact-limited electrical injection current along the $x$ direction from the $s$-subband is \cite{PhysRevLett.121.056802, jensen2017introduction}:
\begin{equation} \label{current}
    \mathcal{J}^{(s)}_{\varphi} = \frac{e}{(2\pi)^2} \int \text{d}^2k v_x^{(s)}(k_x, k_y) \mathcal{T}(k_x) f(\varepsilon_{\mathbf{k},\varphi}, \varepsilon_F, T)
\end{equation}
where $f(\varepsilon_{\mathbf{k},\varphi}, \varepsilon_F, T)$ is the carrier distribution function and $\mathcal{T}( k_x )$ is the transmission function. For thermionic injection, the overbarrier condition of $k_x>\tilde{k}_\text{B}$ is required \cite{PhysRevLett.92.106103}. Here $\Theta(x)$ is the Heaviside step function and $\tilde{k}_\text{B}^{(s)}$ is the $x$-component of the wave vector obtained from Eq. (3) by setting $\varepsilon_{\mathbf{k},\varphi} = \Phi_{B0}$ and $k_y = 0$ , i.e. $\tilde{k}_\text{B}^{(s)} \equiv k_x \left(\varepsilon_{\mathbf{k},\varphi}^{(s)}, k_y\right) = \sqrt{\Phi_{B0}/2\Lambda_+^{(s)}}$ where $\Phi_\text{B}$ is the Schottky barrier height (SBH) measured from zero-energy. 
Assuming that the $\Phi_{B0}$ is sufficiently larger than $\varepsilon_F$, the thermionic charge injection occurs in the carrier distribution function can be reasinably approximated by the semiclassical Maxwell-Boltzmann distribution function. Equation (\ref{current}) can be rewritten as:
\begin{widetext}
\begin{equation}\label{current2}
    \mathcal{J}^{(s)}_{\varphi} = \frac{e}{(2\pi)^2\hbar} \int_{\tilde{k}_\text{B}}^{\infty} \text{d}k_x \int_{-\infty}^{\infty} \text{d}k_y  \left(2\Lambda_+^{(s)}k_x^2  - s\alpha \psi' k_y \right) \exp\left(- \frac{\Lambda_+^{(s)}(\varphi) k_x^2 + \Lambda_-^{(s)}(\varphi) k_y^2 - s\alpha \psi'(\varphi) k_xk_y - \varepsilon_F}{k_bT}\right)
\end{equation}
We express the current into two components, $\mathcal{J}^{(s)}_{\varphi} = \mathcal{J}^{(s)}_{\text{I},\varphi} + \mathcal{J}^{(s)}_{\text{II},\varphi}$ which corresponds to each term of the carrier group velocity [see Eq. (\ref{vg})]. The first term can be rewritten as
\begin{equation}
    \mathcal{J}^{(s)}_{\text{I},\varphi} = \frac{2e(k_BT)^{3/2}}{(2\pi)^2 \hbar \sqrt{\Lambda_-^{(s)}} }  \exp\left( \frac{\varepsilon_F}{k_BT} \right) \int_{\sqrt{\frac{\Phi_{B0}}{ k_BT} }}^{\infty} \bar{k}_x\text{d}\bar{k}_x \exp\left(- \bar{k}_x^2\right) \int_{-\infty}^{\infty} \text{d} \bar{k}_y \exp\left( -\bar{k}_y^2 + A\bar{k}_x\bar{k}_y \right) 
\end{equation}
where $A \equiv s\alpha \psi' / \Lambda_+^{(s)} \Lambda_-^{(s)}$, and we perform the variable substitutions: $\bar{k}_y \equiv \left(\Lambda_+^{(s)} / k_BT \right)^{1/2} k_x$ and $\bar{k}_x \equiv \left(\Lambda_-^{(s)} / k_BT \right)^{1/2} k_y$. Using the standard integral identities $\int_{-\infty}^{\infty} \exp\left(-x+ax^2\right) = \sqrt{\pi}\exp\left(a^2/4\right)$ and $\int_{a}^{\infty} \exp\left(-bx^2\right) = (1/2b)\exp\left(-ab\right)$, we obtain
\begin{equation}\label{J1}
    \mathcal{J}^{(s)}_{\text{I},\varphi} = \frac{\sqrt{\pi}e(k_BT)^{3/2}}{(2\pi)^2 \hbar \sqrt{\Lambda_-^{(s)}} }  \exp\left( \frac{\varepsilon_F}{k_BT} \right) \left(1 - \frac{\alpha^2\psi'^2}{4\Lambda_+^{(s)}\Lambda_-^{(s)}} \right)^{-1}\exp\left[ - \left(1 - \frac{\alpha^2\psi'^2}{4\Lambda_+^{(s)}\Lambda_-^{(s)}}\right)\frac{\Phi_{B0}}{k_BT} \right]
\end{equation}
Using the same substitution of variables $k_x \to \bar{k}_x$ and $k_y \to \bar{k}_y$, the second term of Eq. (\ref{current2}) can be similarly analytically reduced to yield, $\mathcal{J}^{(s)}_{\text{II},\varphi} = - \left(\alpha^2\psi'^2/4\Lambda_+^{(s)}\Lambda_-^{(s)}\right) \mathcal{J}^{(s)}_{\text{I},\varphi}$. The $s$-subband total current is obtained by the summation of $\mathcal{J}^{(s)}_{\text{I},\varphi}$ and $\mathcal{J}^{(s)}_{\text{II},\varphi}$,
\begin{equation}
    \mathcal{J}^{(s)}_\varphi = \frac{e}{ \hbar^2} \left(\frac{m}{\Lambda_-^{(s)} }\right)^{1/2} \left(\frac{k_BT}{2\pi}\right)^{3/2} \exp\left(\frac{\alpha^2\psi'^2}{4\Lambda_+^{(s)}\Lambda_-^{(s)}}\frac{\Phi_{B0}}{k_BT} \right) \exp\left( - \frac{\Phi_{B0}-\varepsilon_F}{k_BT} \right),
\end{equation}
and the \emph{total} thermionic injection current across an altermagnetic Schottky contact is:
\begin{equation}\label{J_AMS}
    \mathcal{J}_{\text{AMS},\varphi} = \frac{em^{1/2}}{\hbar^2}\left(\frac{k_BT }{ 2\pi}\right)^{3/2}\sum_{s}\left( \frac{1}{1 - s\frac{\alpha}{\alpha_c}\psi} \right)^{1/2} \exp\left( \frac{\psi'^2}{1 - \frac{\alpha^2}{\alpha_c^2} \psi^2} \frac{\alpha^2}{4\alpha_c^2}  \frac{\Phi_{B0} }{k_BT}\right) \exp\left(-\frac{\Phi_{B0} - \varepsilon_F}{k_BT}\right)
\end{equation}
When the altermagnetism is switched off ($\alpha = 0$), we recover the 2DEG lateral thermionic emission results of $\mathcal{J}_\text{2D} = (em^{1/2}/\hbar^2) \left(k_BT / 2\pi\right)^{3/2} \exp\left[-(\Phi_{B0} - \varepsilon_F)/k_BT\right]$ \cite{PhysRevLett.121.056802}. 
For $\varphi = (0, \pi/4)$, the angular terms becomes $\psi = (0, 1/2)$ and $\psi'=(-1,0)$, yielding
\begin{subequations}\label{J_final}
    \begin{equation} \label{J0}
        \mathcal{J}_{\text{AMS}, 0} = \frac{2em^{1/2}}{\hbar^2}\left(\frac{k_BT}{2\pi}\right)^{3/2} \exp\left(\frac{\alpha^2}{4\alpha_c^2} \frac{\Phi_{B0}}{k_BT}\right) \exp\left(-\frac{\Phi_{B0} - \varepsilon_F}{k_BT}\right)
    \end{equation}
    \begin{equation} \label{J45}
        \mathcal{J}_{\text{AMS}, \pi/4} = \frac{2em^{1/2}}{\hbar^2}\left[\left(1-\frac{\alpha}{2\alpha_c}\right)^{-1/2}+ \left(1+\frac{\alpha}{2\alpha_c}\right)^{-1/2} \right]\left(\frac{k_BT}{2\pi}\right)^{3/2} \exp\left(-\frac{\Phi_{B0} - \varepsilon_F}{k_BT}\right)
    \end{equation}
\end{subequations}
where the two terms in the square parenthesis of Eq. (\ref{J45}) in the $\varphi = \pi/4$ ($3\pi/4$) case represent the spin-up (spin-down) and spin-down (spin-up) injection currents, respectively, whereas the injection current of the $\varphi = 0, \pi$ cases [Eq. (\ref{J0})] are spin-degenerate. Equation (\ref{J_final}) thus immediately hints at the generation of spin-polarized injection current at $\varphi = \pi/4$, as demonstrated below.
\begin{figure*}[t]
    \centering
    \includegraphics[scale=0.58]{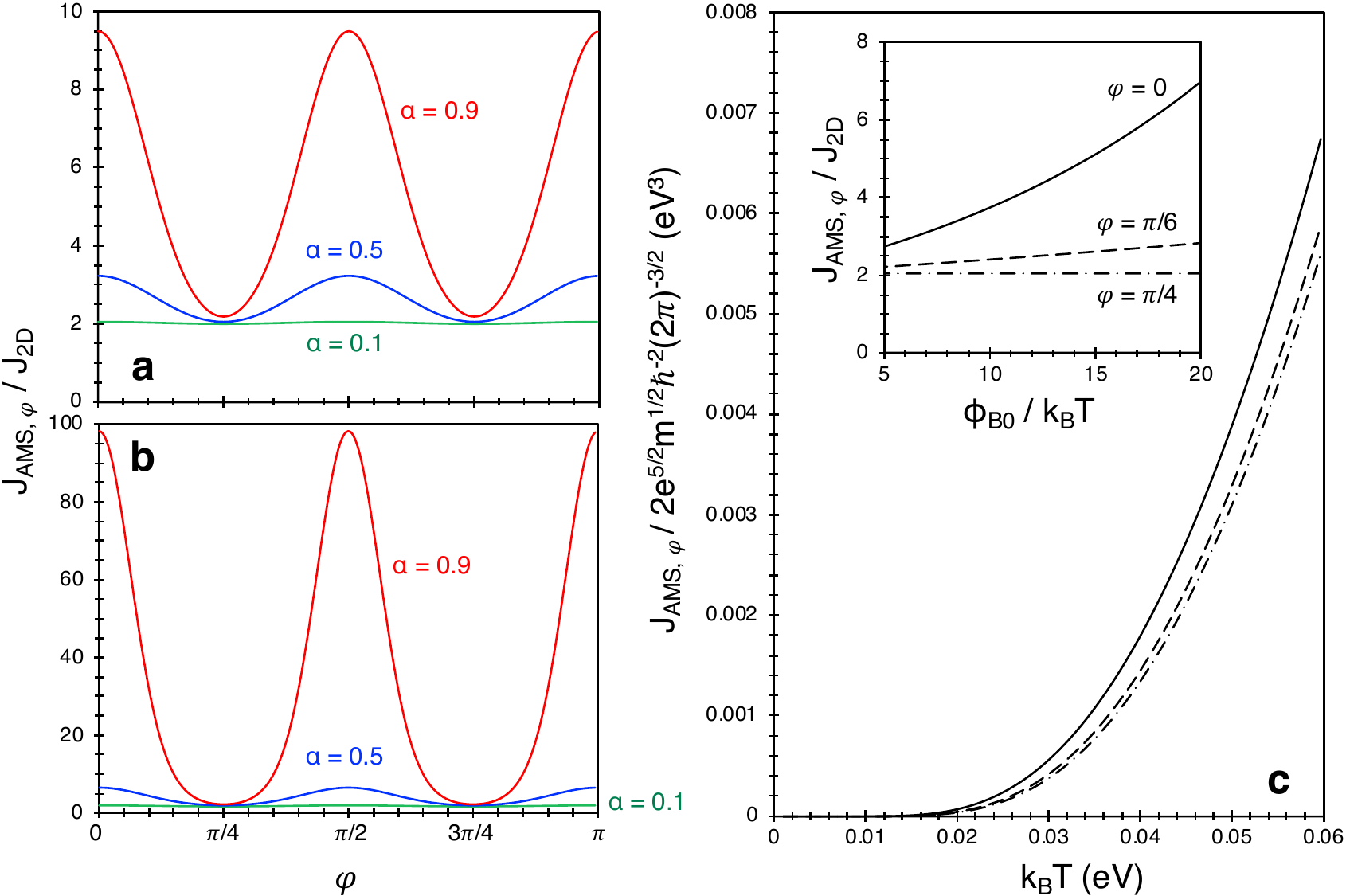}
    \caption{\textbf{Total thermionic injection current across AM/S Schottky contact.} (a) $\varphi$-dependence of the injection current (normalized by 2D injection current) with $\Phi_{B0} = 0.2$ eV and $T=300$ K. (b) same as (a) but for $\Phi_{B0} = 0.5$ eV. (c) Temperature dependence of $\mathcal{J}_{\text{AMS}, \varphi}$. Inset shows the $\Phi_{B0} /k_BT$-dependence of $\mathcal{J}_{\text{AMS},\varphi} / \mathcal{J}_{2D}$}
    \label{fig:2}
\end{figure*}

\end{widetext}

The total injection current exhibits a four-fold rotational symmetry for the orientation axis of the altermagnet as shown in the angular dependence of the normalized current, i.e. $\mathcal{J}_{\text{AMS},\varphi} / \mathcal{J}_\text{2D}$ in Fig. 2(a). Maximum injection current occur at $\varphi = 0, \pi/2, \pi$ when the elliptical Fermi surface lobes of both spins cuts across the $x$-axis at a 45-degree angle [see Fig. \ref{fig:2}(a)]. In contrast, minimum injection current occurs at $\varphi = \pi/4, 3\pi/4$ when the major axis of the Fermi surface lobe of one spin and the minor axis of the Fermi surface lobe of the other spin both align with the $x$-axis. 
The $\varphi$-contrasting spin injection current can be amplified via a larger altermagnetic strength parameter $\alpha$. At $\alpha/\alpha_c = 0.9$, the maximum injection current can be over 40 times larger than the minimum injection current [Fig. \ref{fig:2}(b)], thus suggesting the lattice orientation of the altermagnetic electrode sensitively controls the magnitude injection current across the AM/S contact. 
Such orientation-contrasting injection current can also be amplified by a larger $\Phi_{B0}/k_BT$ as shown in the inset of Fig. \ref{fig:2}(c). Such effect arises from the fact that the $\varphi=0$ injection current acquires an additional exponential term [i.e. first exponential term in Eq. (\ref{J0})] which significantly enhances the $\varphi = 0$ injection current, while such exponential term is completely absent in the $\varphi = \pi/4, 3\pi/4$ case.  
In Fig. \ref{fig:2}(c), the full temperature dependence of $\mathcal{J}_{\text{AMS},\varphi}$, normalized by a factor of $2e^{5/2}m^{1/2}\hbar^{-2} (2\pi)^{-3/2}$, shows an increasing injection current at elevated temperature which is a commonly expected feature of thermionic process. 

\begin{figure}\label{fig:3}
    \includegraphics[scale = 0.858]{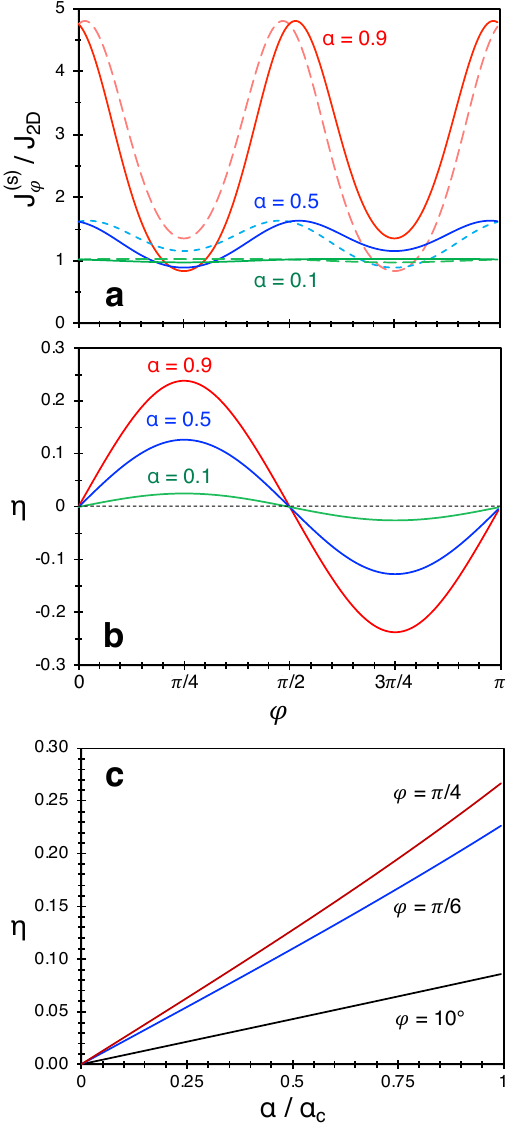}
    \caption{\textbf{Spin-polarized transport in altermagnetic Schottky contact.} (a) Spin-dependent current density (normalized by $J_\text{2D}$ as a function of interface configuration angle $\varphi$. The solid and dahsed curves denote $s=1$ and $s=-1$ spin subband, respectively. Spin polarization efficiency of the injection current as a function of (b) $\varphi$; and of (c) the altermagnetic strength parameter $\alpha/\alpha_c$.}
    \label{fig:3}
\end{figure}

Intriguingly, spin-polarized injection current can be generated in AM/S contact even though the altermagnetic electrode has a net zero magnetization. Such spin-polarized injection originates from the momentum-space spin-splitting of the Fermi surfaces, which generates spin-polarized transport at $\varphi = \pi/4, 3\pi/4$ orientation. Generally, the spin polarization is defined as $\eta \equiv (\mathcal{J}_\varphi^{(+)} - \mathcal{J}_\varphi^{(-)}) / (\mathcal{J}_\varphi^{(+)} + \mathcal{J}_\varphi^{(-)})$. From Eq. (\ref{J_AMS}), the $\eta$ of AM/S contact is obtained as
\begin{equation}\label{pol}
    \eta(\alpha, \varphi) = \frac{\left(1 + \psi(\varphi)\frac{\alpha}{\alpha_c}\right)^{1/2} - \left(1 - \psi(\varphi)\frac{\alpha}{\alpha_c}\right)^{1/2}}{\left(1 + \psi(\varphi)\frac{\alpha}{\alpha_c}\right)^{1/2} + \left(1 - \psi(\varphi)\frac{\alpha}{\alpha_c}\right)^{1/2}}
\end{equation}
which depends only on two parameters: $\varphi$ (through $\psi(\varphi)$ and $\alpha/\alpha_c$. 
At $\varphi = 0$ (or $\pi$), the $\psi = 0$ which yields $\eta = 0$ whereas at $\varphi = \pi/4$ (or $3\pi/4$ for opposite spin), $\psi = 1/2$ and the maximum spin polarization efficiency occurs at $\alpha/\alpha_c = 1$ with maximal efficiency, $\eta_\text{max} = (\sqrt{3/2} - \sqrt{1/2}) / (\sqrt{3/2} + \sqrt{1/2}) \approx 26.8 \%$.

The spin-polarized transport across AM/S contact is illustrated in Fig. \ref{fig:3}. In general, the normalized injection current through one spin subband, i.e. $\mathcal{J}_\varphi^{(s)}/\mathcal{J}_\text{2D}$, does not overlap with that of the opposite spin subband. Spin-polarized transport is thus expected to occur for most crystal orientation of antiferromagnet, except at the special angles of $\varphi = 0, \pi/2, \pi$ [Fig. \ref{fig:3}(a)] where the Fermi surface lobes of \emph{both} spin subbands make an angle intersect the transport direction ($x$-axis) with an angle $\pi/4$, leading to spin-degenerate transport current. 
The spin polarization efficiency, $\eta$ as defined in Eq. (\ref{pol}), exhibits peak polarization at $pi/4$ and $3\pi/4$ for spin-up and spin-down polarized current, respectively.
The spin polarization is enahcned by a larger $\alpha/\alpha_c$ [Figs. \ref{fig:3}(a) and (b)]. This aspect can also been seen in Fig. \ref{fig:3}(c) which suggests the $\eta$ increases almost linearly with $\alpha/\alpha_c$ for three representative altermagnet orientation angles of $\varphi = 10^\circ, \pi/4, \pi/6$. The spin polarization approaches a maximal value of $26.8\%$ as $\alpha/\alpha_c \to 1$. Although such $\eta$ is lower than other ferromagnetic contacts, which can reach well over 30\% \cite{hanbicki2002efficient, osipov2004efficient, low2008modeling, albrecht2003spin}, the possibility of achieving a spin-polarized injection current in a contact with net-zero magnetization and no SOC still suggests an unusual feature not straightforwardly achievable in conventional tunneling structures.
As the antiferromagnetic ordering of altermagnetic metal can be tuned electrically via SOT effect, the AM/S contact can potentially function as an electrically-tunable spin injector \cite{takeuchi2021chiral}.

%
%
%

We now briefly discuss three further extensions based on the AM/S contact injection model developed above. Firstly, the formalism can be generalized into 3D contact by straightforwardly introducing an additional $k_z$-integration. Assuming that the dispersion along the $k_z$ direction is $\varepsilon_z(k_z) = \hbar^2 k_z^2 / 2m_z$ where $m_z$ is the electron effective mass along the $k_z$ direction, the injection current across a 3D AM/S Schottky contact along the $xy$-plane becomes
\begin{eqnarray}\label{J_AMS_3D}
    \mathcal{J}_{\text{AMS},\varphi}^{(\text{3D})} &=& \frac{em^{1/2}m_z^{1/2}k_B^2}{(2\pi)^2\hbar^3}T^2\sum_{s}\left( \frac{1}{1 - s\frac{\alpha}{\alpha_c}\psi} \right)^{1/2} \nonumber \\
    &\times&\exp\left( \frac{\psi'^2}{1 - \frac{\alpha^2}{\alpha_c^2} \psi^2} \frac{\alpha^2}{4\alpha_c^2}  \frac{\Phi_{B0} }{k_BT}\right) \exp\left(-\frac{\Phi_{B0} - \varepsilon_F}{k_BT}\right) \nonumber \\
\end{eqnarray}
which reduces to the Richardson-Dushman law \cite{dushman1930thermionic} in the non-altermagnetic limit of $\alpha = 0$. 
We further note that in the 3D case, spin-polarized injection current is absent for transport along the $z$-axis, since the spin-polarized Fermi surface of the two spins are entirely projected onto the contact interfaces. 
Secondly, the image-force barrier-lowering effect (IFBL) \cite{brahma2023importance} can be included in the AM/S Schottky contact injection model. The image charge effect prodcued by an injected electron leads to the lowering of the Schottky barrier height at the contact. The IFBL has been recently developed for 2D semiconductor under various contact configurations \cite{PhysRevApplied.20.044003}. Consider the contact interface situating at $x=0$ with altermagnetic metal and semiconductor at $x<0$ and $x>0$, respectively. We consider a simple potential barrier profile given by $U(x) = \Phi_{B0} - eVx/L - \beta e^2/16\pi\epsilon x$ where $V$ is the applied bias voltage, $\epsilon$ is the dielectric constant of the dielectric environment, and $\beta$ is a factor dependent on the contact configuration \cite{PhysRevApplied.20.044003} with $\beta = 2/\pi$ for a 2D/2D contact. To determine the barrier lowering term, we identify the maximum point of $U(x)$ via $\left.dU(x)/dx\right|_{x=x_\text{max}} = 0$, which yield the maximum barrier height of $U(x_\text{max}) = \Phi_{B0} - \Delta \Phi_{B0}$ with $\Delta \Phi_{B0} \equiv \sqrt{\beta e^3 V / 4 \pi \epsilon L}$. The IFBL can thus be included in the altermagnetic thermionic injection current in Eqs. (\ref{J_AMS}) and (\ref{J_AMS_3D}) by replacing $\Phi_{B0}$ with $\Phi_{B0} - \Delta \Phi_{B0}$. 
Finally, the Shockley diode equation \cite{shockley1949theory} for AM/S interface can be constructed as $J_{\text{AMS}}(V,T) = \mathcal{J}_{\text{AMS},\varphi}^{(N\text{D})} \left[\exp\left( eV/n k_BT \right) - 1 \right]$ where $n$ is the ideality factor, and $N = 2, 3$ denotes the 2D and 3D reversed saturation thermionic injection current density given by Eqs. (\ref{J_AMS}) and (\ref{J_AMS_3D}), respectively, for 2D and 3D altermagnetic AM/S contacts. 

In conclusion, we develop an analytical theory of thermionic charge injection across an altermagnetic-metal/nonmagnetic semiconductor based on a semiclassical transport formalism. The injection current across such contact exhibits strong interface orientation dependence. Interestingly, spin-polarized injection current can flow across an altermagnetic Schottky contact even when the system has net-zero global magnetization and the SOC effect is completely absent. 
The spin polarization efficiency is ultimately capped at around 26.8\%. 
These findings reveal a surprising potential of altermagnetic metal to produce spin-polarized spin injection into a nonmagnetic semiconductor without involving ferromagnetic ordering. 
The absence of stray field in altermagnetic Schottky contact suggests that altermangetism may offer an alternative route to achieve ultracompact spintronic devices that are immune to stray field, composed of simpler and stable materials (such as RuO$_2$ \cite{PhysRevX.12.040501}) and resilient against external magnetic field. 
We expect the thermionic charge injection model developed in this work to serve as a harbinger of theoretical and experimental studies of contact-limited charge transport in altermagnetic interface. 
Myriads of transport and contact phenomena, such as quantum tunneling injection \cite{PhysRev.102.1464, simmons1963generalized, forbes2019murphy, lepetit2017electronic}, injection in the presence of spin-flip and spin diffusion processes \cite{PhysRevB.62.R4790, tang2015electrical}, photo-induced charge emission \cite{jensen2007general}, contact barrier inhomogeneity \cite{PhysRevApplied.14.054027} and the first-principle simulation of altermagnetic-metal/semiconductor contacts based on density functional theory \cite{wang2021efficient, cao2022designing} and \textit{ab initio} quantum transport simulations \cite{PhysRevApplied.20.014050, guo2022interfacial}, shall form potentially interesting topics to investigate in future works. 

\begin{acknowledgments}
This work is supported by the Singapore Ministry of Education Academic Research Fund Tier 2 (Award No. MOE-T2EP50221-0019) and the SUTD Kickstarter Initiatives (SKI) (Award No. SKI 2021\_01\_12).
\end{acknowledgments}

\section*{AUTHOR DECLARATIONS}

\subsection*{Conflict of Interest}

The authors declare no conflict of interest.

\subsection*{Author Contributions}

\textbf{Yee Sin Ang}: Investigation (lead); Formal analysis (lead); Visualization (lead); Writing – original draft (lead); Conceptualization (lead); Funding acquisition (lead); Formal analysis (equal); Writing – review and editing (lead).

\section*{DATA AVAILABILITY}

The data that support the findings of this study are available from the corresponding author upon reasonable request.


%

\end{document}